# Energy characteristics of multi-muon events in a wide range of zenith angles

A. G. Bogdanov, N. S. Barbashina, D. V. Chernov, L. I. Dushkin, S. S. Khokhlov, V. A. Khomyakov, V. V. Kindin, R. P. Kokoulin, K. G. Kompaniets, A. A. Petrukhin, V. V. Shutenko, I. I. Yashin, E. A. Yurina
*National Research Nuclear University MEPhI (Moscow Engineering Physics Institute), Moscow, 115409, Russia*

G. Mannocchi, G. Trinchero
*Osservatorio Astrofisico di Torino – INAF, 10025, Italy*

O. Saavedra
*Dipartimento di Fisica dell' Universita di Torino, 10125, Italy*

Change of the energy characteristics of muon bundles with an increase of the primary cosmic ray particles energy can be a key to solving the problem of muon excess in the extensive air showers (EAS) observed in a number of experiments. In this work the data on the energy deposit of multi-muon events in a wide range of zenith angles (and as a consequence in a wide range of primary particles energies) obtained with NEVOD-DECOR setup over a long time period are presented. The experimental data are compared with the results of simulations of EAS muon component performed using CORSIKA code.

## 1. INTRODUCTION

Investigations of the energy characteristics of EAS muon component are important to solve the "muon puzzle" – the problem of muon excess (in comparison with calculations carried out in the framework of the existing hadron interaction models even for extremely heavy cosmic ray mass composition) observed in several experiments on detection of ultrahigh energy cosmic rays [1-3]. One of possible approaches to the solution of this task is the measurement of muon bundle energy deposit in the detector material. Such experiment is currently being conducted at the NEVOD-DECOR setup located in the MEPhI.

It is well known that muon energy loss in the matter almost linearly depends on their energy: $dE/dX \sim a + bE$, where $a$ is ionization loss, $b$ is energy loss for radiation processes (bremsstrahlung, direct production of $e^+e^-$ pairs and photonuclear interaction). Therefore, a significant deviation of the measured dependence of the muon bundle energy deposit on the primary cosmic ray particle energy from the expected one, calculated for conventional muon generation mechanisms, can indicate to the change of nucleus-nucleus interactions, e.g., due to the inclusion of new physical processes [4].

## 2. EXPERIMENTAL SETUP AND DATA

NEVOD-DECOR setup includes two main detectors: Cherenkov water calorimeter NEVOD [5] with volume 2000 $m^3$ and coordinate-tracking detector DECOR [6] with area 70 $m^2$. Detecting system of the NEVOD setup (Figure 1) is a spatial lattice of 91 quasi-spherical measuring modules (QSMs) arranged in 25 vertical strings (9 strings with 3 QSMs and 16 strings with 4 QSMs). The distances between the modules are 2.5 m along the detector and 2 m across it and over the depth. Each QSM consists of 6 low-noise 12-dynode FEU-200 photomultipliers with flat 15 cm diameter photocathodes directed along rectangular coordinate axes, that permits to detect Cherenkov radiation from any direction with practically the same efficiency. The total number of PMTs is 546. A wide dynamic range (1 – $10^5$ photoelectrons) is provided due to 2-dynode signal readout and allows to measure energy deposit of muon bundles.

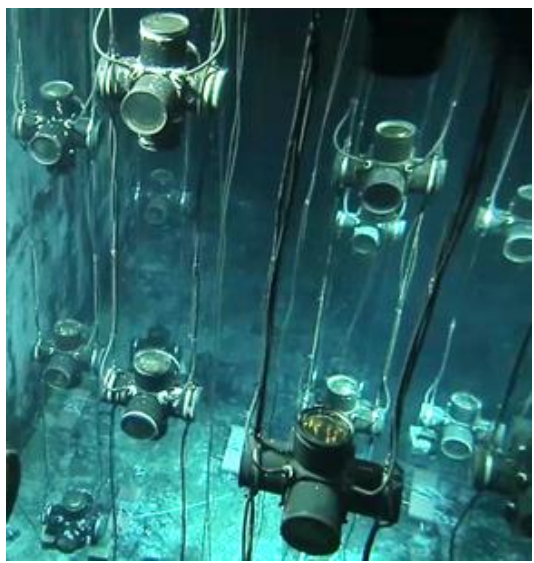

Figure 1. Detecting system of Cherenkov water calorimeter NEVOD.

The DECOR setup consists of 8 vertical supermodules (SMs) placed in the galleries of the experimental building around the water tank of the NEVOD detector (Figure 2). Effective area of one SM is 8.4 $m^2$. Each SM is 8-layer assembly of plastic streamer tube chambers with resistive cathode coating. Earlier these chambers were used in the NUSEX experiment. The chamber is a basic element of the detector, it contains 16 tubes with inner cross section $9×9$ $mm^2$, the length of the chamber is 3.5 m. A three component gas mixture (Ar + $CO_2$ +

n-pentane) provides chamber operation in a limited streamer mode. The layers (planes) of the chambers are equipped with a two-dimensional system of external readout strips: 256 X- and 256 Y-channels in each plane. The total number of registration channels is 32768. Angular and spatial accuracy of the reconstruction of muon tracks crossing the SM is better than 1° and 1 cm, respectively.

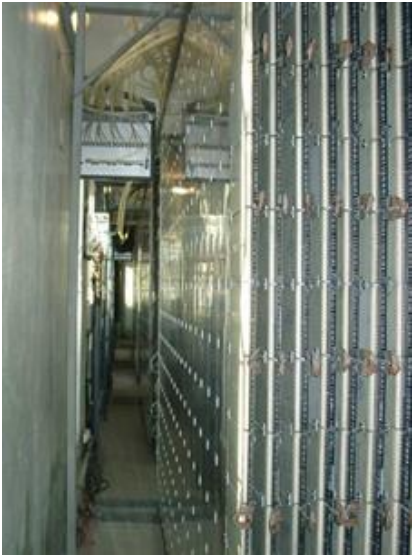

Figure 2. Supermodules of coordinate-tracking detector DECOR.

In this work, the results of the analysis of the data on the energy deposit of inclined muon bundles accumulated during 3 long-term measurement series from May 2012 to May 2016 ("live" time is equal to 23 216 hours) are presented.

From these experimental data 39542 events with muon multiplicity $m \geq 5$ and zenith angles $\theta \geq 55°$ were selected; in addition 15084 muon bundles with smaller zenith angles $40° \leq \theta < 55°$ were sampled. In order to improve the identification of muon tracks, the events were selected in two 60°-wide azimuth angle sectors where 6 of 8 DECOR SMs are screened by the NEVOD detector water tank. In this case, the mean threshold energy of muons is about 2 GeV. Selection procedure of multi-muon events included several steps: trigger level (3-fold coincidence of signals from different SMs within 250 ns time gate); software reconstruction and selection; final event classification and counting of tracks by several operators. An example of a muon bundle event detected in NEVOD-DECOR setup is shown in Figure 3.

The local density of muons in the events of $D$ is obtained according to data of the detector DECOR taking into account the bias of the estimate due to Poisson fluctuations in the number of particles hitting detector and the steep slope of muon density spectrum as: $D = (m - \beta)/S_{\text{det}}$, where m is muon multiplicity, $\beta \approx 2.1$ is the integral slope of density spectrum, $S_{\text{det}}$ is the effective area of DECOR SMs for given direction of muon bundle arrival. A new approach to the study of the EAS – a method of local muon density spectra (LMDS) [7] allows to estimate the typical energy of primary cosmic ray particles, which give the main contribution to events selected by muon density, according to the DECOR data (multiplicity of muons and zenith angle). The energy deposit of muon bundles is measured in the Cherenkov water calorimeter NEVOD. As a first approximation, the sum of the all PMTs signals $\Sigma N_{\text{pe}}$ (photoelectrons) is proportional to the local muon density $D$ [8], therefore in the further analysis the specific energy deposit $\Sigma N_{\text{pe}}/D$ is considered.

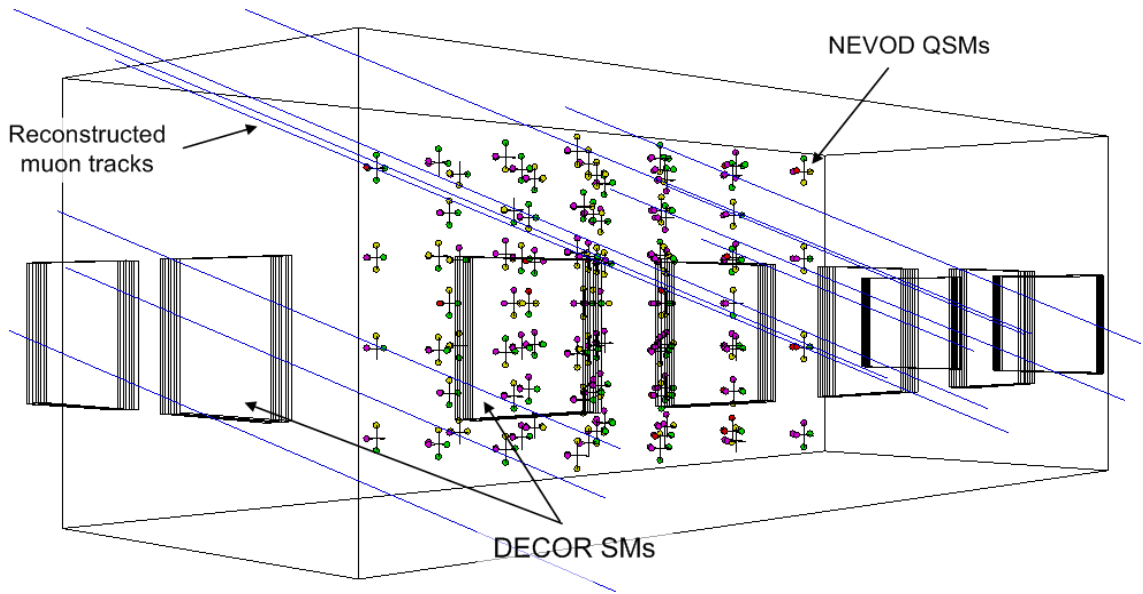

Figure 3. An example of muon bundle event detected in the NEVOD-DECOR setup (thin lines represent reconstruction of muon tracks from DECOR data; small circles are hit phototubes in the NEVOD calorimeter; big rectangles are supermodules of the DECOR).

The local density of muons in the events of $D$ is obtained according to data of the detector DECOR taking into account the bias of the estimate due to Poisson fluctuations in the number of particles hitting detector and the steep slope of muon density spectrum as: $D = (m - \beta)/S_{det}$, where $m$ is muon multiplicity, $\beta \approx 2.1$ is the integral slope of density spectrum, $S_{det}$ is the effective area of DECOR SMs for given direction of muon bundle arrival. A new approach to the study of the EAS – a method of local muon density spectra (LMDS) [7] allows to estimate the typical energy of primary cosmic ray particles, which give the main contribution to events selected by muon density, according to the DECOR data (multiplicity of muons and zenith angle). The energy deposit of muon bundles is measured in the Cherenkov water calorimeter NEVOD. As a first approximation, the sum of the all PMTs signals $\Sigma N_{pe}$ (photoelectrons) is proportional to the local muon density [8], therefore in the further analysis the specific energy deposit $\Sigma N_{pe}/D$ is considered.

## 3. RESULTS AND DISCUSSION

In Figure 4, the dependence of the muon bundle average specific energy deposit $\Sigma N_{pe}/D$ on zenith angle $\theta$ is shown. At moderate angles (40-55°) the exponential decrease of the energy deposit is due to residual contribution of the electron-photon and hadron components of EAS to the NEVOD detector response [9]. For zenith angles > 55°, where almost only muons remain in the events, the increase of the average specific energy deposit with a rise of zenith angle is observed, that indicates the increase of the average muon energy in the bundles. Curves in the figure represent the expected dependences of the energy deposit of muon bundles on zenith angle in assumption that the mass composition of primary cosmic rays is only protons or pure iron nuclei. Arrows in the figure show the estimates of typical energies of the primary cosmic rays. The detection of muon bundles of various multiplicities in a wide range of zenith angles allows to explore the interval of primary particle energies from $10^{16}$ to $10^{18}$ eV.

Calculations are based on simulations of EAS muon component by means of the CORSIKA code [10] (version 7.5) for primary protons and iron nuclei with energies in a range of $10^{14} - 10^{19}$ eV. The simulation was performed taking into account the Earth's magnetic field, as a model of hadronic interactions SIBYLL-2.3 was used. Average energies of muons comprising the bundles for different zenith angles were obtained in accordance with the LMDS method [7]; muon energy loss in water (ionization and radiation) were calculated by interpolation of tabulated data [11]. Finally, the theoretical curves were normalized to the experimental data at zenith angle of 60°. As seen from the figure the experimental data are in a good agreement with the results of calculations. Estimates show that the average muon energy in the bundles grows from 150 GeV at zenith angle 55° to 500 GeV at 85°, that is about 3-5 times more than the average muon energy in the

EAS. Therefore, muons in the bundles are closely associated with the most energetic hadrons of the EAS, that gives the possibility of testing existing models of nucleus-nucleus interactions at ultrahigh energies.

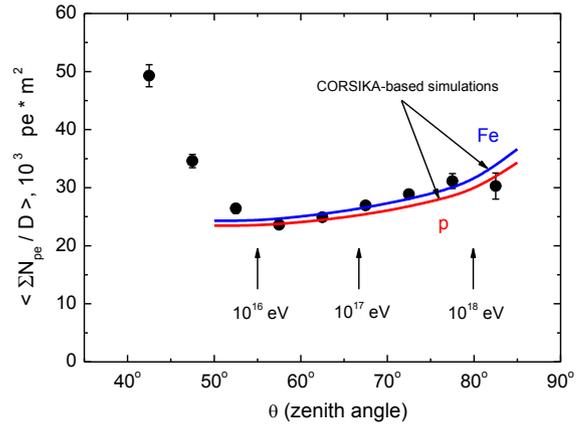

Figure 4. Dependence of the muon bundle average specific energy deposit on zenith angle. Points are experimental data; curves represent the expected dependence based on the simulations with CORSIKA for primary protons and iron nuclei; arrows indicate typical energies of primary cosmic ray particles.

Dependences of the average specific energy deposit of muon bundles in the NEVOD detector on muon density $D$ for two intervals of zenith angles: 55-65° and 65-75° are presented in Figure 5; last three data points (in order of density increase) contain 469, 182, 49 and 81, 26, 9 events, respectively. In fact, these dependences can be considered as the change of the specific energy deposit of muon bundles with the increase of primary particle energy (see the arrows in the figure). The expected dependences of the energy deposit of muons bundles in EAS formed by the primary protons and iron nuclei on the muon density are shown by curves; the same normalization as for Figure 4 was applied here. From the Figure 5 it is clear that the results of the calculations show a decrease of muon bundle energy deposit (and, accordingly, the average muon energies) with an increase of the primary particle energies. At the same time, the experimental data indicate to a possible increase of muon bundle energy deposit at densities $D > 1$ particles/m², i.e. at primary particle energies $E_0 > 10^{17}$ eV. Such densities (1-2 particles/m²) correspond to about 30-60 tracks in the detector DECOR (5-10 tracks per SM), which can be easily identified and counted. It is important to note that for two considered intervals of zenith angles the typical energies of primary particles that give contribution to events with the same muon density differ by 3-4 times. Thus, the range of primary cosmic rays energies covered in this experiment is about $10^{16} - 10^{18}$ eV.

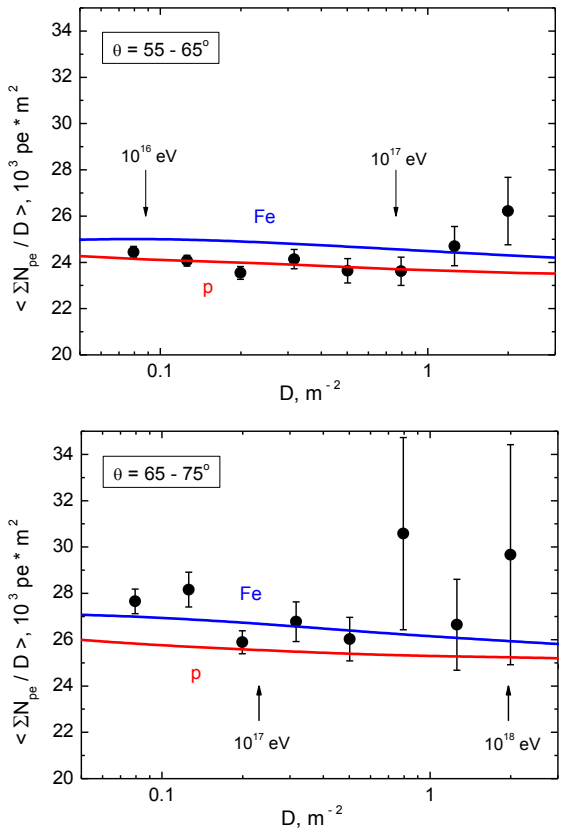

Figure 5. Dependence of average specific energy deposit on muon density for different zenith angles (up, for zenith angles 55-65°, and down, for zenith angles 65-75°; points are experimental data; curves represent the expected dependence based on the simulations with CORSIKA for primary protons and iron nuclei; arrows indicate typical energies of primary cosmic ray particles).

## 4. CONCLUSION

The experiment on the investigation of the energy deposit of inclined muon bundles formed as a result of interactions of primary cosmic ray particles with energies $10^{16} - 10^{18}$ eV is being conducted at the NEVOD-DECOR setup. The measurements of zenith-angular dependence of the average specific energy deposit in the Cherenkov water detector are in a reasonable agreement with CORSIKA-based simulations of EAS muon component and confirm the increase of the average energy of muons in the bundles at large zenith angles. A some indication to an increase of the average specific energy deposit compared to the expectation at primary particle energies above $10^{17}$ eV has been found.

## Acknowledgments

This work was performed at the Unique Scientific Facility "Experimental complex NEVOD" with the financial support of the Ministry of Education and Science of the Russian Federation (project RFMEFI59114X0002, MEPhI Academic Excellence Project and government task). Simulations were carried out using the resources of the MEPhI high-performance computing center.